\documentclass{aip-cp}

\usepackage[numbers]{natbib}
\usepackage{rotating}
\usepackage{graphicx}
\usepackage{tikz}
\usepackage{textcomp}
\usepackage{placeins}
\newcommand{\defineblack}{\definecolor{c}{rgb}{0,0,0}}
\newcommand{\aj}{Astronomical Journal}

\begin{document}

\title{A Pointing Solution for the Medium Size Telescopes for the Cherenkov Telescope Array}

\author[aff1]{D. Tiziani\corref{cor1}}
\author[aff2]{M. Garczarczyk}
\author[aff3]{L. Oakes}
\author[aff3]{U. Schwanke}
\author[aff1]{C. van Eldik}
\author{the CTA Consortium}
\eaddress{www.cta-observatory.org}

\affil[aff1]{Friedrich-Alexander-Universit\"at Erlangen-Nu\"urnberg, Erlangen Centre for Astroparticle Physics, Erwin-Rommel-Str. 1, D 91058 Erlangen, Germany}
\affil[aff2]{DESY, D-15738 Zeuthen, Germany}
\affil[aff3]{Institut f\"ur Physik, Humboldt-Universit\"at zu Berlin, Newtonstr. 15, D 12489 Berlin, Germany}

\corresp[cor1]{Corresponding author: domenico.tiziani@fau.de}

\maketitle

\begin{abstract}
An important aspect of the calibration of the Cherenkov Telescope Array is the pointing, which enables an exact alignment of each telescope and therefore allows to transform a position in the sky to a point in the plane of the Cherenkov camera and vice versa. The favoured approach for the pointing calibration of the medium size telescopes (MST) is the installation of an optical CCD-camera in the dish of the telescope that captures the position of the Cherenkov camera and of the stars in the night sky simultaneously during data taking. The adaption of this approach is presented in this proceeding.
\end{abstract}

\section{INTRODUCTION}
The Cherenkov Telescope Array (CTA) is the next generation gamma-ray observatory which is presently in a pre-construction phase. As an array of Imaging Atmospheric Cherenkov Telescopes (IACT) it makes use of the fact that very-high-energy (VHE) photons induce particle showers in the Earth's atmosphere. The shower particles move faster than the local speed of light and emit Cherenkov light. This light can be collected by large mirror dishes and imaged to a dedicated Cherenkov camera.

The pointing of an IACT describes the exact direction the telescope is pointed to and therefore also the transformation of a point in the image plane of the Cherenkov camera to a position in the sky. Precise knowledge of the pointing of each telescope in an array is important for the calculation of the original direction of an air shower. In order to determine the pointing, one does not solely rely on the mechanics and the tracking system of the telescope, which can be affected by imprecision. Some mechanical effects, like bending of telescope components by gravity or wind loads, can lead to mis-pointing, i.e. the real pointing differs from the expected pointing. The pointing of IACTs is typically measured using optical starlight. In this proceeding, the adaption of a method for pointing calibrations of the medium size telescopes of CTA is described.

\section{SINGLE-CCD CONCEPT}
\begin{figure}[h]
\begin{tabular}{cc}
\includegraphics[width=0.45\textwidth]{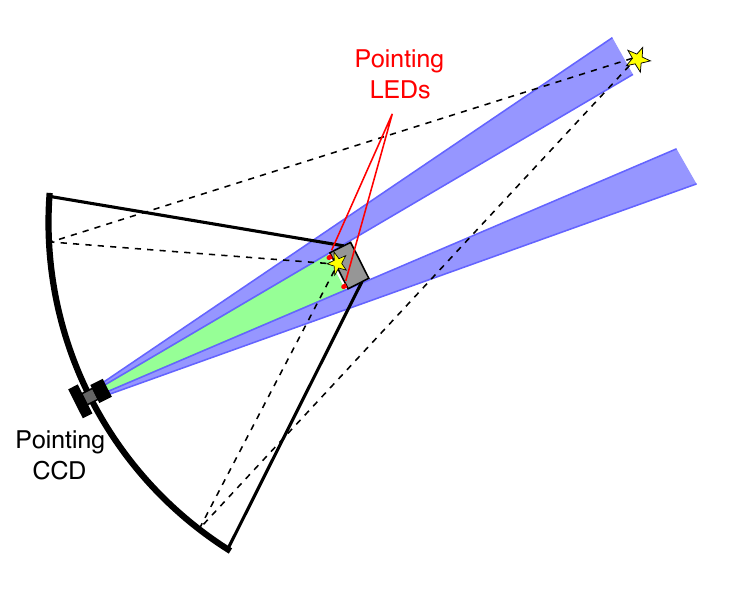} &\includegraphics[width=0.55\textwidth]{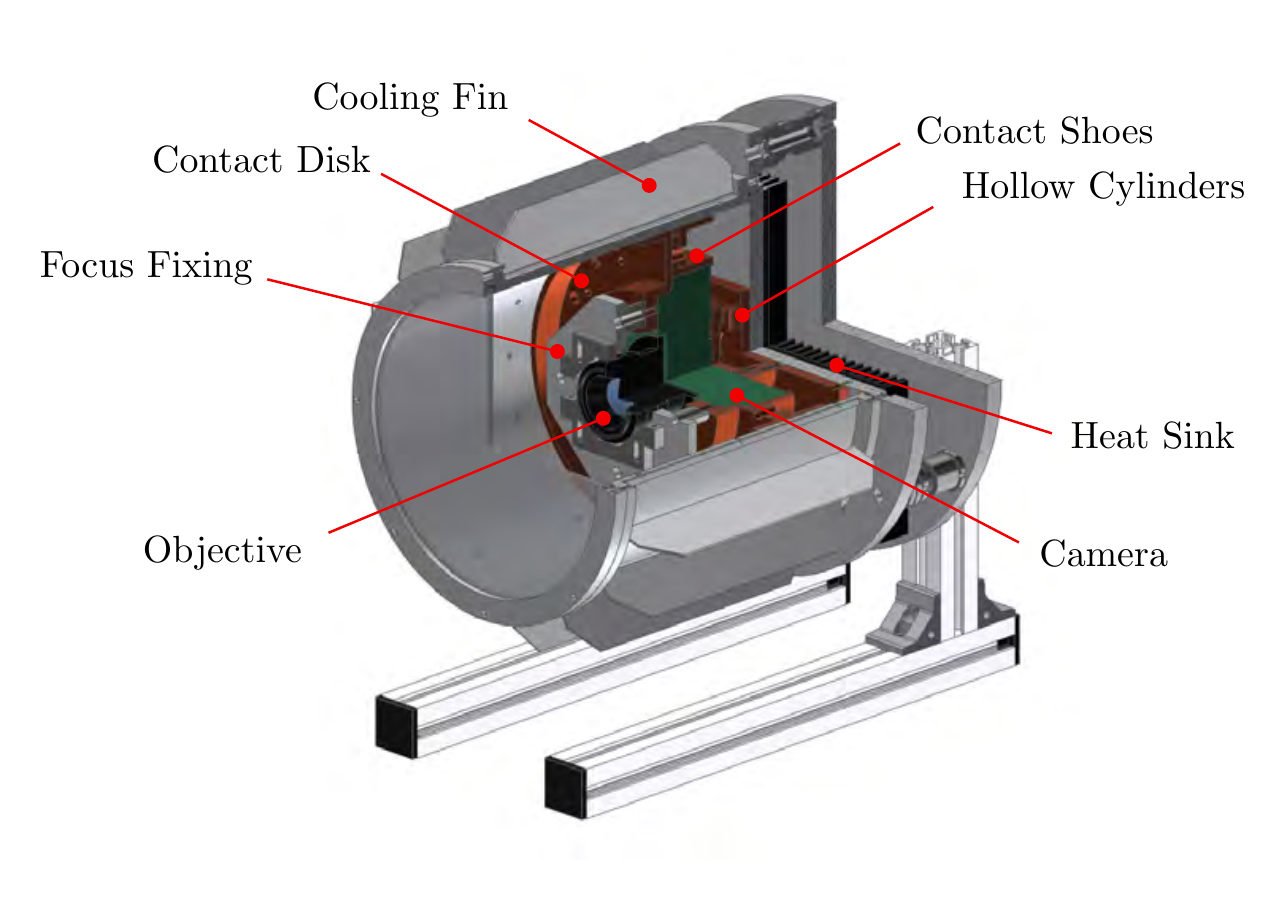}\\
(a)&(b)
\end{tabular}
  \caption{(a): Schematic view of a Cherenkov telescope with a single Pointing CCD camera mounted to the mirror dish. Both pointing LEDs on the Cherenkov camera and stars surrounding it can be imaged. (b): Schematic view of the proposed CCD-camera in its casing. The rigid support structure (focus fixing) and heat conducting elements (contact disk, contact shoes, cooling fins, heat sink, and hollow cylinder) are highlighted.}
  \label{fig:pointing}
\end{figure}

The concept that is presented here uses one optical CCD-camera with a large field-of-view that is mounted in a central position of the mirror dish of an IACT. This camera observes both several LEDs on the Cherenkov camera and the night sky surrounding the latter (see Figure \ref{fig:pointing} (a)). With the captured images the position of the Cherenkov camera relative to the dish and the alignment of the telescope can be determined. This concept has already been tested successfully by the H.E.S.S. collaboration \citep{lennarz}.

Another concept that is currently being investigated uses two separate cameras for these two tasks. The advantages of the Single-CCD concept over this solution are, besides lower costs and complexity, reduced systematics from transformations from one camera to another camera with the downside of reduced angular resolution. 

\subsection{Camera Hardware}
The pictures are taken with an Apogee ASPEN CG8050 (electronic shutter, chip dimension 3296 $\times$ 2472 pixels, pixel size 5.5\,\textmu m) astronomical camera. A 50\,mm f/1.8 Nikkor lens allows for images with a per-pixel resolution of 22'' on the sky and a field-of-view of 20.5 $\times$ 15.5\,deg$^2$. The camera provides active chip cooling via a Peltier element. With this technique, chip expansions during observations can be avoided which would be a critical systematic uncertainty in the precise determination of the pointing. The camera can be read out via a standard Ethernet interface.

\begin{figure}[h]
  \centerline{\resizebox{0.70\textwidth}{!}{\input{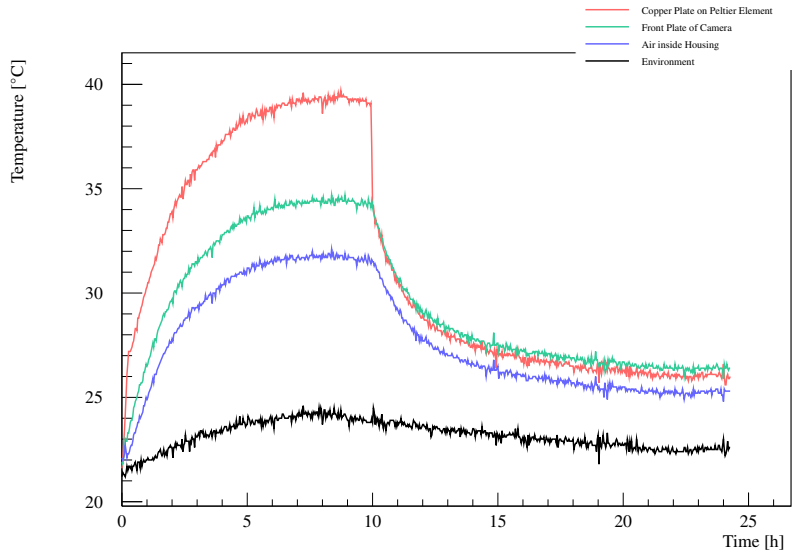}}}
  \caption{Measurement of temperatures at different locations in the camera housing. At time t = 0\,h, when all parts of the structure have environmental temperature, camera and chip cooling are switched on. At 10\,h camera returns to stand by. After the heating-up phase, the air temperature inside the casing stays well below 40$^\circ$C (the maximum operating temperature of the camera) at an environmental temperature of $\sim 25^\circ$C. (the maximum operating temperature defined for CTA).}
  \label{fig:temperature}
\end{figure}

The CCD-camera is enclosed in a custom-made casing that complies to the IP67 standard (see Figure \ref{fig:pointing} (b)). This housing is rigidly attached to the center of the mirror dish to prevent movement of the camera relative to the dish. The camera is also mounted inside the housing in a mechanically stiff way. The casing comprises several heat conducting elements made from solid aluminum that transfer the heat power of the Peltier element and camera electronics to the outside surface and the surrounding air. With this passive cooling technique the air temperature inside the casing can be kept at a level that allows a save operation of the camera. This has also been tested in a laboratory setup (see Figure \ref{fig:temperature}). Long term tests of the hardware under more realistic conditions are currently performed at the MST prototype in Berlin.

\subsection{Pointing Reconstruction and Tests with Simulated Images}
Both simulated and real images (taken at the MST prototype telescope) are used to verify the usability of the Single-CCD-Concept for the pointing of the MSTs.

In order to determine the pointing, the CCD-images are analyzed with the open source software \mbox{\textit{astrometry.net}} \citep{2010AJ....139.1782L}. This software searches for stars in a given image and identifies constellations with the help of pre-compiled index files. With this information it sets up transformations between pixel and celestial coordinates. 

\begin{figure}[h]
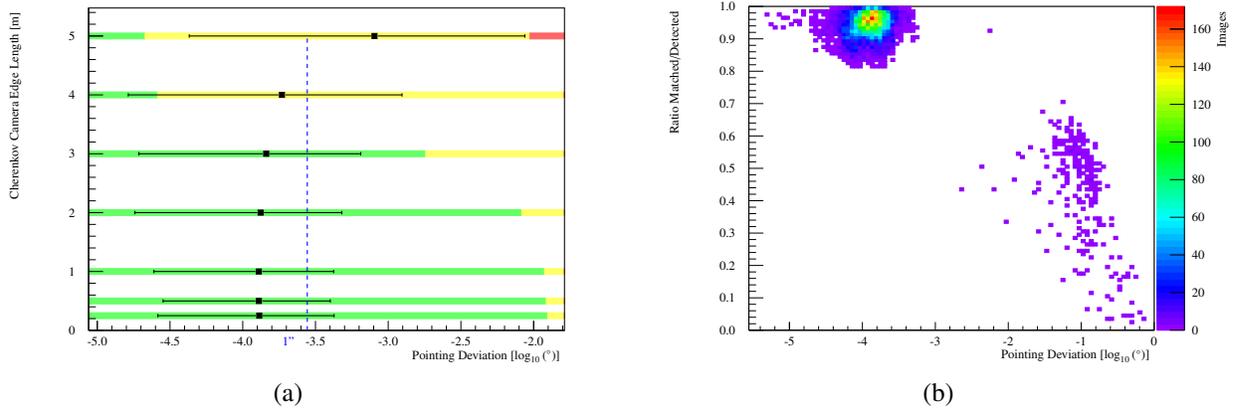

\begin{tabular}{cc}
\resizebox{0.50\textwidth}{!}{\input{graph_cam.tex}} & \resizebox{0.50\textwidth}{!}{\input{corr_no_cam.tex}}\\
(a) & (b)
\end{tabular}
  \caption{(a): Study of the expected precision of pointing reconstructions from simulated data as a function of Cherenkov camera size. The plot has been produced with 10000 simulated images for each camera size. Black data points represent the median precision, black error bars the 99\% containment interval. The green bars indicate the fraction of images that could be solved correctly by \textit{astrometry.net} and were used for the reconstruction. The dashed blue line highlights a precision of 1''. Bigger camera sizes lead to an increase of the fraction of bad reconstructions (yellow bars). With a quality cut on the number of identified stars w.r.t. the number of detected stars in an image, these can be sorted out. This separation criterion is shown in (b) for images without an obstructing Cherenkov camera.}
  \label{fig:size}
\end{figure}

Since the Cherenkov camera of the telescope will block the view to the central part of the sky in the pointing images, the dependence of the precision of the pointing reconstruction on the size of the Cherenkov camera was tested with simulated data (see Figure \ref{fig:size}). For this purpose, a simulation software was developed which allows for the creation of pointing images from a virtual MST. The investigations show that the precision is well below 1'' for a Cherenkov camera with an edge length of 3\,m, which is about the size of the cameras that will be deployed at the telescope. 

\begin{figure}[h]
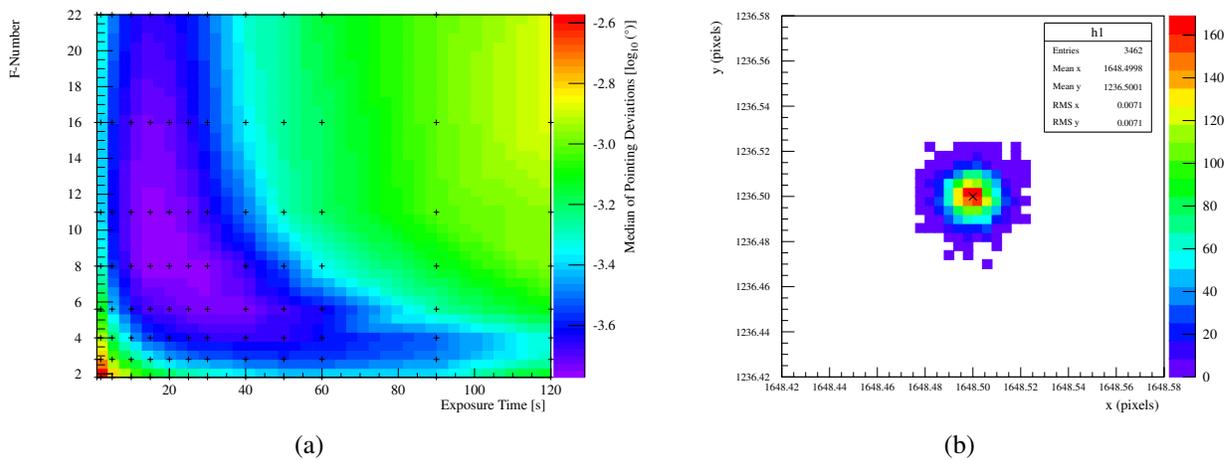

\begin{tabular}{cc}
\resizebox{0.56\textwidth}{!}{\input{graph_expAp_medians.tex}} & \resizebox{0.44\textwidth}{!}{\input{camera_center.tex}}\\
(a) & (b)
\end{tabular}
  \caption{(a): Median precision of pointing reconstruction for different simulated configurations of aperture and exposure time. Black pluses indicate simulated configurations. The colored histogram is the result of an interpolation. Apertures of f/8 to f/11 together with exposure times in the range of 10-24\,s allow for most precise pointing reconstructions representing an optimum of large photon statistics and small rotation of the field-of-view and spot blurring. (b): Reconstructed position of the Cherenkov camera center for a set of 3462 simulated images. The center was calculated from the positions of 8 symmetrically positioned LEDs in the corners of the Cherenkov camera plane. Positions are given in units of CCD-camera pixels; each pixel translates to about 22'' on the sky. The simulated center of the camera is at $(x,y) =  (1648.5, 1236.5)$ (marked with an x).}
  \label{fig:expAp}
\end{figure}

The simulation was also used to find an optimal balance between aperture and exposure time. On the one hand small apertures and short exposure times reduce effects of spot blurring and rotation of the field-of-view during the exposure, respectively. On the other hand the star spots are also fainter and more difficult to detect, with fewer detected stars leading to reduced precision. The assessment suggests apertures in the range of f/8 and f/11 and exposure times between 10\,s and 25\,s (see Figure \ref{fig:expAp} (a)).
Figure \ref{fig:expAp} (b) shows the expected precision with which the position of the Cherenkov camera in an image can be determined during pointing measurements. The statistical uncertainty is about $7 \times 10^{-3}$ CCD pixels for a simulated aperture of f/8. This translates to a precision of below 0.2'' per axis i.e. $\sim$ 10\,\textmu m per axis in the camera plane.

\section{Tests at the MST Prototype}

\begin{figure}[h!]
\begin{tabular}{cc}
\includegraphics[width=0.5\textwidth]{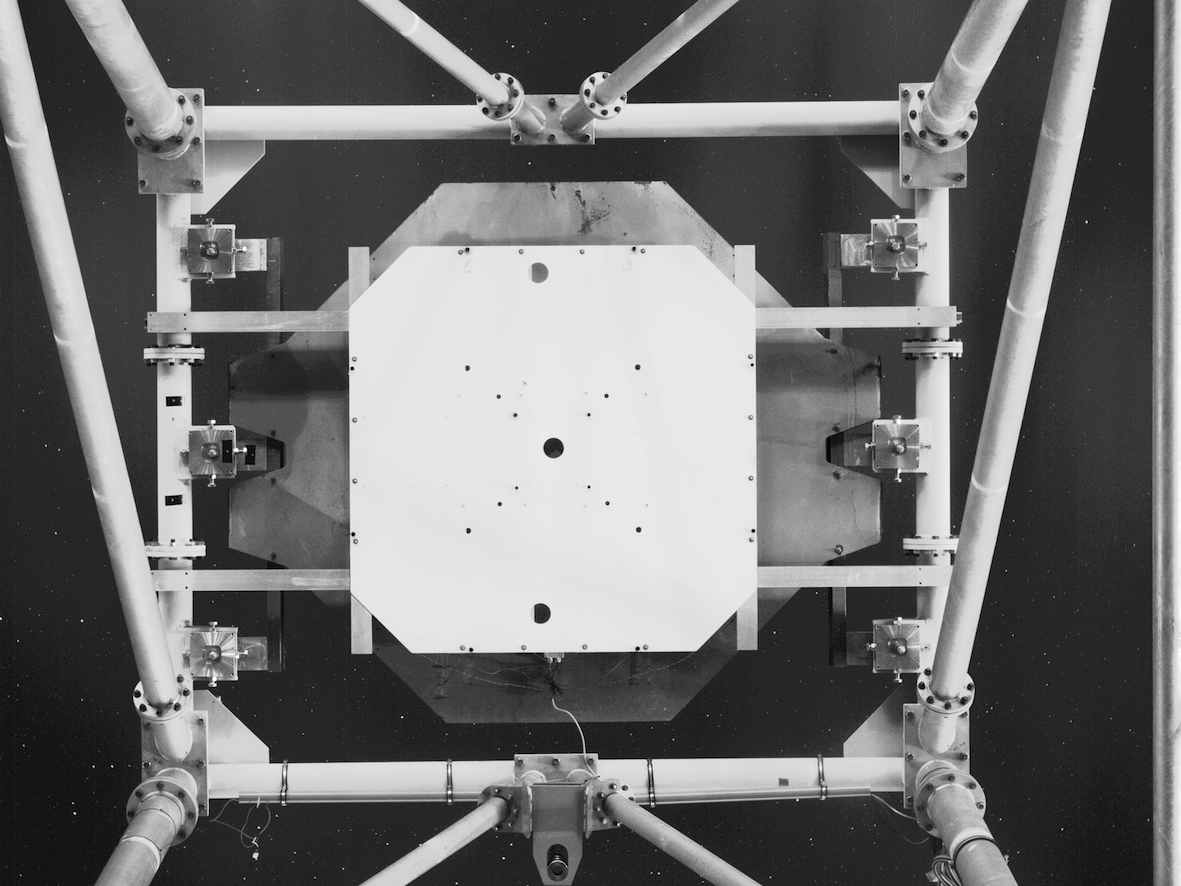} & \includegraphics[width=0.5\textwidth]{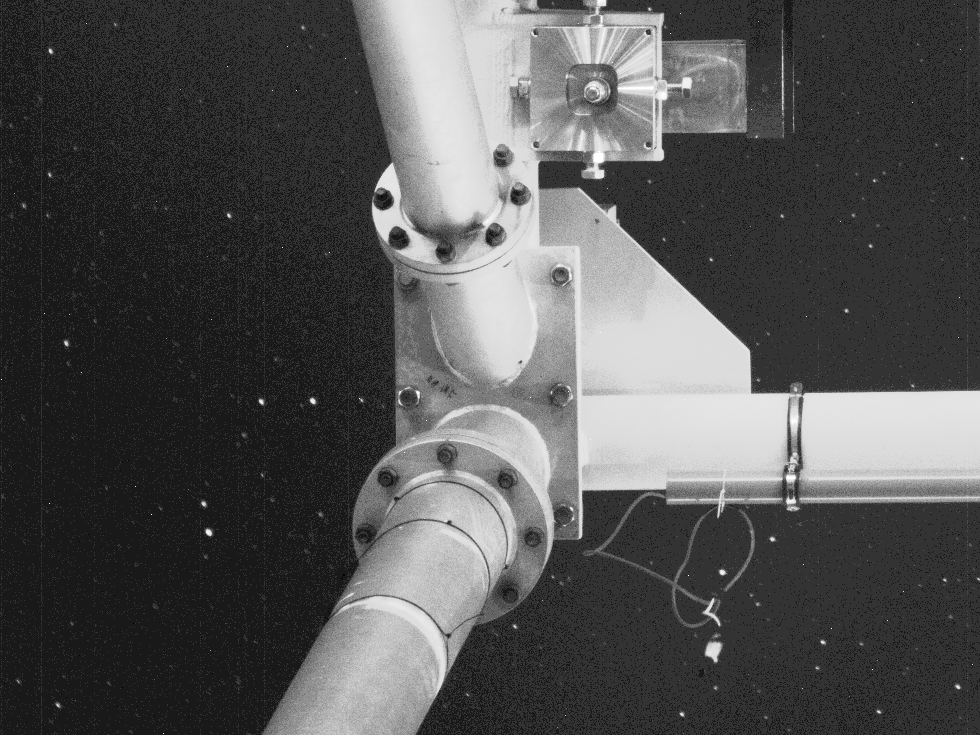}\\
(a) & (b)
\end{tabular}
  \caption{Image taken with the Single-CCD-camera mounted on the MST prototype. The image (a) shows the whole image with a dummy Cherenkov camera in the center and its support structure around it in front of the dark night sky. In (b) the lower left part of the image is magnified. Stars up to magnitudes of about 7 can be identified. The image was taken with an aperture of f/8 and an exposure time of 20\,s.}
  \label{fig:sky}
\end{figure}
Checks of the hardware and software are currently performed at the MST prototype telescope which is located in Berlin/Adlershof. Images during first tracking tests show that $>70$ stars up to magnitudes of $\sim 7$ can be detected in a single exposure in the rather bright night sky of Berlin. Figure \ref{fig:sky} shows such an image. It is clearly visible that a large part of the sky is obscured by the telescope structure. The impact of this blocking on the precision of pointing reconstructions is currently under investigation using real data from the prototype telescope. 

\section{CONCLUSION}
Results from simulations and test runs at the prototype telescope show that the Single-CCD-Concept is a promising solution for pointing measurements of the medium size telescopes of CTA. It has been proven by simulations that the pointing of such a telescope can be reconstructed to a precision of better than 1'' by applying this method. Ongoing performance tests are needed to promote this concept to a state in which it can be used at the operational gamma-ray observatory.
\FloatBarrier
\section{ACKNOWLEDGMENTS}
We gratefully acknowledge support from the agencies and organizations under Funding Agencies at www.cta-observatory.org. This work was supported by the German Ministry of Education and Research under grant identifier 05A14WE2.

\nocite{*}
\bibliographystyle{aipnum-cp}


\end{document}